\documentclass[preprint]{aastex}

\slugcomment{To appear in AJ.}

\shorttitle{Quantitative Morphology of Intermediate-Redshift Galaxies}
\shortauthors{Trujillo et al.}

\begin{document}

\title{Quantitative Morphology of the Intermediate-Redshift Galaxy Cluster
 Abell 2443 from Ground-Based Imaging -- Evidence for a galaxy concentration index correlation with cluster
density}

\author{I. Trujillo}
\affil{Instituto de Astrof\'{\i}sica de Canarias, E-38200, La Laguna, Tenerife, Spain}
\email{itc@ll.iac.es}
\author{J.A.L. Aguerri}
\affil{Astronomische Institut der Universit\"at Basel, Venusstrasse 7, CH-4102 Binningen, Switzerland}

\and

\author{{C.M. Guti\'errez} and {J. Cepa}}
\affil{Instituto de Astrof\'{\i}sica de Canarias, E-38200, La Laguna, Tenerife, Spain}

\begin{abstract} 
We present broad-band photometry and provide a quantitative analysis of
the structure of galaxies in the inner region of the Abell Cluster 2443
($z \sim 0.1$). The galaxy parameters have been derived by fitting a
two-component model (S\'ersic $r^{1/n}$ bulge and exponential disk)  to
a magnitude-limited sample. Using a new method of analysis which takes
into account the effects of seeing on the structural parameters and
 considers the ellipticity as an active parameter, we avoid systematic
errors arising from assumptions of circular symmetry. 76\% of the
sample galaxies  were classified with these models, the rest were
morphologically peculiar. For the spiral galaxies, the relation between $n$
and $B/D$ is consistent with the trend observed in nearby field galaxy
samples.  The S\'ersic index $n$ (which can be considered as a 
concentration index) of the elliptical galaxies is correlated with the
local surface density of the cluster. Monte Carlo simulations were used
to check the reliability of the method and determine the magnitude
selection criteria.

 \end{abstract}


\section{Introduction}

Following the discovery of
the Butcher \& Oemler (1978, 1984) effect considerable observational effort has been devoted to understanding the change in the
properties of the galaxies in high density environments. Only recently, thanks to the high
spatial resolution imaging achieved with the $Hubble$ $Space $ $Telescope$
$(HST)$ and the improvement in the ground-based observations, it has been
established that the morphological properties of galaxies in rich clusters at
intermediate redshift differ dramatically from those in nearby clusters. There
is an increase in the number of spiral population, a factor of  2--3 less S0
galaxies, and the fraction of ellipticals is already as larger as, or larger than, in
nearby clusters. Most of the observational work has been developed in the range
z=0.3--0.5 (Couch et al. 1994, 1998; Dressler et al. 1994; Wirth, Koo, \& Kron
1994; Dressler et al. 1997; Oemler, Dressler, \& Butcher 1997; Smail et al.
1997).  Few studies have been done in the z $\sim$ 0.1--0.2 regime (Fasano et
al. 2000) which is a crucial range for a better understanding of  galaxy evolution in dense environments.

The claimed morphological evolution of galaxies in distant clusters is based,
mostly, on visual classifications of these objects.  This classification system
is based on the Hubble scheme, in which the ratio of the spheroidal to the
disk components is a key parameter  differentiating between elliptical and
spiral galaxies.  Visual  classifications  were developed on the basis of
nearby, bright normal  galaxies (Sandage 1961; Sandage \& Tammann 1987), and 
are founded on the assumption that  the structure of galaxies follows  certain
archetypes. But, the ability to classify visually by Hubble type is
increasingly difficult for faint and/or high-redshift galaxies, and it is
therefore necessary to use a quantitative profile decomposition method to
retrieve the physical properties of the observed  galaxies.
Quantitative classification has two major advantages over visual
classification: (1) it is reproducible, and (2) biases can be understood and
carefully characterized through simulations that are treated as real data. The
Hubble deep field (HDF) is a good example where, for the same sample of
galaxies, different classification schemes have been applied (visual: van den
Bergh et al.\ 1996, non-parametric: Abraham et al.\ 1996 and profile
decomposition: Marleau \& Simard (1998)) obtaining substantially different
results.

The use of quantitative morphology enables the recovery of reliable information on
the structural parameters (shape, size, axial ratios, etc.) of the galaxies.
The information contained in the structural parameters play an important role
that is  required to understand the evolution and origin of 
galaxies. Visual morphology classification  must be considered only as the
first step in the detailed studies of the properties of the galaxies in high
density environments. Furthermore, galaxies  in dense regions are
undergoing tidal friction, high-speed impulse encounters and mergers. These
processes can affect  the density profile of the galaxies and must be compared
with those from field galaxies.

Automatic classification systems are based on the
fitting of structural models to  the surface-brightness profiles of galaxies.
In this case it is assumed that the surface-brightness profile of every
structural component follows some analytical law. Nowadays, the spheroidal
component is usually modeled with a S\'ersic profile (S\'ersic 1968) and the
disk component with  an exponential law.  An alternative scheme is based on
non-parametric model-independent classification in which
galaxies are  classified using some quantitative index such as color,
concentration,  and asymmetry of the galactic light distribution (e.g.,
Bershady, Jangren, \& Conselice 2000). 


Seeing scatters light from  objects, producing a loss of resolution in the
images. Morphological classification based on visual inspection from
ground-based images is therefore compromised as redshift increases. For the
same reason, classification systems based on concentration and
asymmetry are also degraded with redshift and need be corrected for the effects
of seeing. Methods  based on fitting structural components to
the surface-brightness profile of the galaxies can be corrected
(Schade et al. 1996).

 In a previous paper (Trujillo et al. 2001a), we  studied the  effects of the
seeing on S\'ersic profiles. We quantified how ignoring these effects on the
observed surface brightness distribution results in significantly different
structural parameters, and consequently different dynamical properties 
inferred from these parameters.
In order to recover the seeing--free structural parameters of distant objects,
we presented the mathematical basis for an accurate
description  of the seeing effects on  surface brightness distributions
following $r^{1/n}$ laws. 

In this paper, we  develope a fitting routine  to estimate the structural
properties of galaxies from the inner region of the intermediate-redshift
($z$=0.103) rich cluster Abell Cluster 2443 following the mathematical analysis
developed in our previous paper. The characteristics of this cluster (richness
and distance) provide an ideal data set for the study of the structural
parameters in an insufficiently probed regime of $z$.  Section 2 describes the
photometric decomposition algorithm. In \S 3 we present the results of
simulations that test the reliability of the method for  measuring galactic
structural parameters from ground-based images of a simulated galaxy sample.
Section 4 describes the results of the photometric decomposition technique
applied to the galaxies in the inner region of Abell Cluster 2443. 

\section{The Fitting Algorithm}

The typical scale size of galaxies studied at intermediate  redshift is
comparable to the seeing obtained with ground-based telescopes (barring
adaptative/active optics). It is therefore absolutely necessary to consider
the effects of seeing on the images, and  very accurate
convolutions of the point spread function (PSF) and the model profiles of
the galaxies are required. Most current fitting algorithms have two main
disadvantages: a) the surface brightness is assumed to be  circularly
symmetric, and b) the convolution procedure is  executed using fast Fourier
transforms. In case a) the  ellipticity parameter is measured usually for the outer most isophotes as a parameter external to the  fit. This procedure ignores the fact that the seeing distorted
surface--brightness distribution is a function  of the seeing and the intrinsic
ellipticity of the object. In case b), even if the elliptical symmetry of the
objects is taken into account, the fast Fourier transform is  not appropriate
when there are strong changes in the  spatial scale of the galaxy profile. Most
of the  profiles in common use (e.g. de Vaucouleurs, S\'ersic) have such shapes. The
inner parts of  the galaxy profiles are steep, and this demands a very
accurate  measure of the high frequencies in the Fourier domain. Consequently,
flatter distributions are obtained as a result of a convolution using the fast 
Fourier transform compared with the traditional methods integrating in the
real  domain.   

The uncertainty in the estimation of the structural parameters also depends on
the signal-to-noise ratio. Decreasing signal-to-noise implies a greater
degeneracy in the possible parameters that fit the same profile. So, it is
important to establish a limiting magnitude where the recovery of parameters is
reliable. This acquires special significance  in the bulge + disk
decomposition. Insufficiently deep images might mask  the disk component of
some galaxies. When a bulge and disk decomposition is tried on these objects a
systematically $lower$ fractional bulge-to-total luminosity ($B/T$) ratio is
obtained. This is because  the disk component of the fit, due to the absence of
points in the outer parts of the galaxy, increases its slope to fit  the
transition region between the bulge and disk; overestimated disks are expected
(see the next section). This problem can also be present in nearby samples (e.g.,
Prieto et al.\ 1992).

To minimize the above mentioned problems, we have developed a 
tool that takes into account the effects of seeing on elliptical 
structures. The mathematical basis of the algorithm is described by 
Trujillo et al.\ (2001a). Briefly, the fitting procedure works using 
 elliptical coordinates ($\xi,\theta$) defined as:
\begin{eqnarray}
x&=&\xi \cos \theta
\nonumber\\ 
y&=&\xi (1-\epsilon)\sin \theta
\end{eqnarray} 
where a different ellipticity, $\epsilon$, is used for each component.

Ellipses are fitted to the objects using the ELLIPSE task from the IRAF
package.  The surface-brightness profile along the semi-major axis (the $\xi$
axis) and  the isophotal ellipticity profile are fitted simultaneously. The
ellipticity profile is uniquely determined for any given seeing and set of
structural parameters. The intrinsic 
ellipticity  is assumed to be constant for all the isophotes.

The goal in correcting the observed profiles for the effects of seeing is to
recover the true intrinsic galaxy light profile. Normally, one does this by
assuming an initial model that represents the intrinsic galaxy profile and then
convolving this with the PSF, iterating until convergence. However, if the
galaxy has elliptical symmetry not only are the inner isophotes circularized,
but the outer isophotes may also be distorted by seeing. More importantly,
although the convolution of a model which is assumed to be spherically
symmetric will fit the observed data, recovered parameters will not represent
the true intrinsic galaxy profile. This is because the convolution of an
elliptical 2D model is different to the convolution of a spherical 2D model. We
have therefore performed the convolution of elliptical two-dimensional models
following  Trujillo et al (2001a). Not doing this can result in systematic
errors.

Two different kinds of PSFs are available in our fitting procedure: Gaussian
and Moffat-type (Moffat 1969). These PSFs are convolved with our surface
brightness profile models. A Levenberg--Marquardt non-linear fitting algorithm 
(Press et al.\ 1992) was used to determine the parameter set which
minimizes $\chi^2$. Extensive Monte-Carlo simulations were done to check the
method and estimate the uncertainties in the recovered parameters. In
particular, the simulations allow us (see Section 3) to gain an understanding
of the limiting magnitude where the procedure is able to obtain reliable
results.

\subsection{The two-dimensional Surface-Brightness Profile Model}

If the intermediate-redshift galaxy population has a roughly similar
distribution of morphological properties as the nearby population, then
structures such as bars, dust lanes, spiral structures, galaxy cores, etc., 
can be neglected due to the small angle scale subtended by these structures at
these distances.  On the other hand, it is expected that any significant
evolution in the galaxy morphology at these redshift ranges will leave its  imprint
in the form of anomalous values of the parameters, or an excess of peculiars
galaxies that can not be parameterized.

The 2D fitted galaxy model has two components (an $r^{1/n}$ bulge and an
exponential disk) with a total of seven parameters: the bulge effective
intensity, $I_{\rm e}$, the bulge effective radius, $r_{\rm e}$, the S\'ersic
index, $n$, the bulge ellipticity, $e_{\rm b}$ ($e \equiv 1-b/a$, $b \equiv$
semi-minor axis, $a \equiv$ semi-major axis), the disk central intensity,
$I_0$, the exponential disk scale length, $h$, and the disk ellipticity,
$e_{\rm d}$. The 2D bulge component is a S\'ersic (1968) profile of the form:

\begin{equation}
 I(\xi)=I_{\rm e}10^{-b[(\xi/r_{\rm e})^{(1/n)}-1]}.
\end{equation} 

The parameter $b$ is set equal to $0.868n-0.142$, so that
$r_{\rm e}$ remains the projected radius enclosing half of the light in this
component  (Capaccioli 1989). The de Vaucouleurs profile is obtained when
$n=4$.  The second, ``disk'', component is a simple exponential  profile of the
form:
 \begin{equation} 
 I(\xi)=I_0e^{(\xi/h)}. 
 \end{equation} 
 The presence of a
``disk'' component does not imply the presence of a rotational disk since many
virially supported systems also have simple exponential  profiles.

\section{Monte-Carlo Simulations}

We have performed Monte-Carlo simulations to test the reliability of our
method.  First, we  tested the ability to recover parameters from 
bulge-only (i.e.\ purely elliptical) structures, and then we  explored
the possibility of carrying out accurate bulge + disk decompositions. In both
cases  we created 150 artificial galaxies with structural parameters randomly
distributed in the  following ranges:

\begin{description} 

 \item a) for  bulge-only structures: 16 $\leq I \leq$ 20, 0$\arcsec$.44  $\leq
r_{\rm e} \leq$ 1$\arcsec$.65, 0.5 $\leq n \leq$ 6, and 0 $\leq \epsilon \leq$
0.6 (the lower limit on $n$ is due to the physical restrictions pointed out in
Trujillo et al. 2001a); 

\item  b) for  bulge + disk structures: 16 $\leq I \leq$ 20,  0$\arcsec$.2
$\leq r_{\rm e} \leq$ 0$\arcsec$.88, 0.5 $\leq n \leq$ 4,  0 $\leq \epsilon_b
\leq$ 0.4, 0$\arcsec$.55 $\leq h \leq$ 1$\arcsec$.65,  0 $\leq B/T \leq$ 0.65,
and 0 $\leq \epsilon_{\rm d} \leq$ 0.6.  \end{description} 

No correlation
between the input structural parameters was imposed. We  also studied
the possibility that our algorithm introduces an artificial correlation between
the recovered parameters by applying a 2D Kolmogorov-Smirnov test (Fasano
\& Franceschini 1987) between the input random data distribution ($n$, r$_e$)
and the output one. The significance level of this analysis was 0.27\footnote{
When the significance level is $>$ 0.20 the two data sets are not significantly
different} indicating that the output distribution was also random. So, no artificial correlations between the output parameters was found.

The artificial galaxies were created by using the IRAF task MKOBJECT. To simulate the real conditions of our observations of the Abell Cluster A2443
(see next section), we added a background sky image (free of sources) taken from
our  $I$-band image (see description below); the  dispersion in the sky
determination was 0.1\%. The seeing in the simulation
was set at 0.88$\arcsec$ (FWHM), as in our observations. The PSF was assumed
to be Gaussian and known exactly. The pixel  scale for all simulations was
0$\arcsec$.11 pixel$^{-1}$ to match our prime-focus  imaging. We follow the same procedure to process both the simulated and actual
data.

\subsection{Bulge-Only Structures}

Figure 1 shows the relative error between the input and output parameters. This
figure also shows the mean relative errors as a function  of the magnitude. We
have divided our simulations into bins of 0.5 mag and computed for each bin the
mean relative error between the input and output magnitudes, $r_{\rm e}$, and
$n$. We have also obtained the mean differences between the input and output
ellipticity. The total number  of simulations (150) resulted in an average
number   of about 20 objects per bin.  All the parameters can be obtained with
an relative error less than 10\% for objects brighter than I magnitude  19.25
(see Table 1). No systematic errors were found for objects brighter than this
magnitude, which can be considered a strict limiting magnitude for the
determination of the structural parameters.

The limiting magnitude is of course a function of the signal-to-noise ratio but
not in the normal manner. Better signal-to-noise permits us to obtain reliable
parameters from  fainter galaxies. From simulations we found that an
increase by a factor 2 in  the signal-to-noise allowed us to determine the
structural parameters for objects two  magnitudes fainter. This is because the
local slope of the surface brightness profiles is a decreasing function of the
radius, so the outermost regions of the profiles are flatter than the inner
ones. Having fixed a detection threshold (i.e. I=19), which depends on the signal-to-noise
factor, there is a rapid increase in the number of points available to fit
the profiles in the outermost regions as the signal-to-noise factor increases.

The difference between the recovered S\'ersic index, which defines the type of
profile, and the input value is shown in  Figure 2 as a function of $n$ . There
is no obvious bias in the  $n$ parameter recovered from the simulations. Most
studies of this kind systematically give lower values of output $n$ for larger
input $n$ (e.g., Marleu \& Simard 1998). The reason for this is the assumption
of circular symmetry in the profile models. This can bias the analysis, and
hence conclusions of studies performed at different redshifts. For instance,
higher redshift galaxies will appear smaller, and are consequently more
affected by the PSF. Not correcting the ellipticity results in the derivation
of smaller values of $n$ for these systems, and therefore effects claims of
structural evolution. We have avoided this problem by taking into account the
ellipticity of the objects in our fitting routine.

\subsection{Bulge + Disk Structures}

The recovery of the real bulge and disk structural parameters is strongly
dependent on the signal-to-noise and the $B/T$ ratio of the objects. In cases
of low signal-to-noise the disk component can be masked by the noise. This can
produce an  over-estimation of the disk, as explained before, which
simultaneously affects  the estimation of the bulge parameters  (e.g., Schade
et al.\ 1996). Sufficiently bright disks do not have this problem, but
intermediate and large $B/T$ galaxies suffer when there is insufficient
signal-to-noise (or equivalent low total magnitudes of the objects). This bias
may complicate the nature of the trend between decreasing $B/T$ and redshift
reported by Marleau \& Simard (1998). From this consideration, it is clearly
necessary to establish a limiting magnitude where the estimation of the
parameters are reliable.

Figure 3 shows the relative error between the input and output parameters, and 
the mean relative errors, as a function of the magnitude.  As in \S 3.1, the
simulations were divided into 0.5 magnitude bins. For each bin, the mean
relative error between input and output bulge and disk magnitudes, $r_{\rm e}$,
$n$, and $h$ was computed. The mean differences between the input and output
ellipticities for the bulge and disk structures were also evaluated. Apart 
from the effective radius, all the parameters can be obtained to an accuracy of
better than 10\% for galaxies with an I-band magnitude brighter than 18.75 (see
Table 2).

Figure 4 shows $d(B/T)\equiv$$B/T$(measured)-$B/T$(input) and mean $d(B/T)$
versus $B/T$(input)  and scale length. We divided the simulations into 0.1 bins
$(B/T)$, the mean differences in $d(B/T)$ going from $B/T=0 $ to $B/T=0.65$
were: 0.08  $(\sigma=0.15)$, 0.02 (0.09), 0.03 (0.09), 0.03 (0.08), 0.02
(0.19), --0.02 (0.16), and 0.05 (0.19). We distinguish between points
brighter and fainter than 18.75 magnitudes. Our results establish that the
$B/T$ relation (which is one of  the most important parameters for classifying
the galaxies) and the scale-length h can be  recovered well  to the limiting
magnitude. As expected, smaller disks (i.e.\ a bigger $B/T$ ratio)
have the greater dispersions in the $B/T$ recoveries. This result is
dependent on the total magnitude of the object (or equivalently,
the signal-to-noise). It is important to stress that the results discussed here
for bulges, or bulge + disk structures, are conservative (errors less than
10\%).

\section{The Abell Cluster 2443 (\boldmath $z=0.103$) }

 Abell 2443 is a rich southern cluster
(richness class 2 and Bautz--Morgan type II) at  $z=0.103$. Its brightest
cluster member, PGC 068859 (identified in our study as ID  = 1), was first
cited in the VLA survey of rich clusters  (Slee, Perley, \& Siegman 1989). 

\subsection{Observations and Data Reduction}

Observations were obtained on August 19, 1998, at the 2.5 m Nordic Optical
Telescope (NOT)  at the Observatorio del Roque de los Muchachos on La Palma.
The prime focus CCD camera was used with pixel scale of 0$\arcsec$.11
pixel$^{-1}$ and the seeing ranged from FWHM 0$\arcsec$.88 to 1$\arcsec$. The
field was observed through $B$, $V$, and $I$ filters with equal integration times (3
$\times$ 900 s) and in the $R$ filter (3 $\times$ 600 s). The mean seeing was
0$\arcsec$.88 (FWHM) in the $I$ images (which were used to investigate the
morphology).
The observed field ($2'.5 \times 2'.5$) covered the central part of the
cluster, which was the inner 300 kpc assuming throughout that $H_0=75$ km s$^{-1}$
Mpc$^{-1}$.

We applied the standard data reduction procedures: bias subtraction, flat
fielding correction, flux calibration using Landolt (1992) standards,
addition of images of the same filter, and sky subtraction. Figure 5 shows
the reduced and calibrated image of the cluster in the $I$ band. The large
extended object at the center is the cD galaxy PGC 068859.

We used the SExtractor galaxy photometry package (Bertin \& Arnouts 1996,
version 2.1.4)  on the images. This  package  is optimized to detect and
measure sources from astronomical images. Basically, SExtractor detects all
sources from an image  above a LOWTHRESHOLD, and which have an area greater
than a MINAREA.  These two parameters are given as input to the program. For 
our extractions we have  taken LOWTHRESHOLD = 1.5 $\sigma$  and MINAREA = 4
pixels, where $\sigma$ is the standard deviation of the sky background of the
images. To get rid of stars we have only studied  extended sources.

A cross-correlation search between the $R$ and $I$ images  enabled us  to 
identify 250 extended objects to $I=23$ mag. From these, we selected 
121 objects that were identified in all four bands. Broad-band aperture
photometry of these objects was done using SExtractor in all the above
mentioned filters (see Table 3). The magnitudes were computed using a variable
aperture for each object. The radius of this aperture was 2.5 $r_{\rm Kron}$,
where $r_{\rm Kron}$ is the Kron  radius  (see Kron 1980). This radius is
obtained from the luminosity profile of each individual object and  is different
for each object.  From our simulations, the  threshold magnitude for an
accurate morphological structure analysis is $I=19$ mag.  This  criteria left us
with 33 objects.

\subsection{Results and Discussion}

Two differents kinds of model fitting were performed: a) a
two-dimensional bulge  + disk decomposition, and b) a two-dimensional
fitting of a S\'ersic law (i.e.\ one structural component). Both models
were fitted for all 33 galaxies selected.  A similar study was carried
out using the $R$-band filter but no significant deviations were found.
We found eight (24\%) galaxies  that could not be fitted by our routine
(ID: 2, 5, 8, 35, 40, 49, 77, and 90). We hereafter refer to these
objects as ``peculiar'' galaxies  since many of them exhibit peculiar
visual morphologies. This left a total  of 25 galaxies. Figure 6 shows
the I-band images of each, along with the radial profile and
ellipticity, and the residuals from the fits. 

Galaxies which satisfy $B/T>0.6$ (19/33) are referred as ``ellipticals''  and
their parameters are presented in Table 5. For these galaxies a better fit to
the profile was found using only a S\'ersic component model. We use the term
``elliptical'' to refer to galaxies that are better fitted with one (S\'ersic law)
component, without necessarily impling a virially supported
system. Those galaxies with $B/T<0.6$ (6/33) are listed in Table 4. The cD
galaxy (ID: 1) was classified by our algorithm as a galaxy with $B/T<0.6$. cD
galaxies are not ``standard'' ellipticals, in fact, these galaxies present
envelopes which may be detected as a separate component in $r^{1/4}$ plots
(Graham et al. 1995). Our algorithm tries to fit the presence of this `extra'
light in the outer parts of the galaxy by adding an extra component to the
profile. By fitting only the inner 5 arcsec with a pure S\'ersic profile, the
best fit is achieved for $n$=1.96. This technique was also used by Graham et al
(1995) and similar values of $n$ are given in their sample. Because of the
``special'' characteristics of these kind of galaxies we have decided not to
include them in the group of ``ellipticals''.

As in most similar studies, contamination from foreground and background
objects is a source of uncertainty. In Figure 7 we plot $(B-I)$ versus
$I$ magnitudes. This figure shows two clearly differentiated branches, a red
one, $(B-I)>3$, and another in which the bulk of the galaxies have $(B-I)<3$. 
We have been able to model five galaxies in the redder branch. These five
galaxies (ID: 26, 29, 85, 86, and 129) have all been labeled as ``ellipticals''
and have high values of $n$ ($n > 4.8$). Also, three of these galaxies (29, 85,
and 86) are near our I=19 cut off limit. It is certainly possible that these
galaxies are field (background) galaxies with a disk component (e.g., Saglia et
al.\ 1997) that we are not able to resolve. Therefore, to err on the side of
caution, we assumed that $(B-I)>3$ galaxies may be background galaxies and  
are not taken into consideration hereafter. Galaxies belonging to the cluster,
$(B-I)<3$, exhibit reasonable values of their parameters with the exception of
object 6. Visual inspection suggests that this object might be an edge on
spiral galaxy that our algorithm mis-classifies. We also exclude this
object  hereafter. Object 11 might be a disk galaxy too but it presents a
reasonable fit and values of the recovered parameters, so in order to be
consistent with our procedure we have maintained it as an elliptical object.

Quantitative morphology permits us to investigate  correlations between the
different structural parameters with those obtained  for nearby galaxy samples.
In Figure 8 we plot $n$ versus the logarithm of the bulge-to-disk luminosity
ratio, $\log(B/D)$, for the galaxies with $B/T<0.6$. There is a clear
correlation between these two quantities, as exists for local field galaxies
(Andredakis, Peletier, \& Balcells 1995; Graham 2001).  There are not enough
galaxies to investigate any significant departure (i.e.\ evolution) from the
results of these authors. The Freeman (1970) relation is also shown in Figure
8. Our results are in good agreement with this relation (solid line). To
evaluate the absolute magnitude of these galaxies we assumed that the galaxies
belong to the cluster and the $K$-correction was implemented following
Poggianti (1997).

Figure 9 shows the relation between the S\'ersic indexes $n$ and a)  the
(projected) distance from the brightest central cluster member (PGC 068859) and
b) the local cluster surface  density (Dressler 1980). We present separately
the values of $n$ for the bulge + disk galaxies and the $n$ values from the
``elliptical'' galaxies.  The values of $n$ for the ``spiral galaxy'' bulges
are, on average, smaller than those obtained for the ``ellipticals''. To
measure the local cluster surface density, the ten nearest (projected)
neighbors to each of the analyzed galaxies were found, after computing the area
involved and correcting for the field galaxy density\footnote{To correct for
field galaxy contamination, galaxies with $(B-I) > 3$ were not taken into
account when computing the density.} the local surface density (galaxies
Mpc$^{-2}$) was evaluated down to $I=23$. The data reveal that elliptical
galaxies with higher values of $n$ are located in the inner part of the
cluster. The Pearson's $r_p$ correlation coefficient between $n$ and central
distance is $r_p$=-0.45; if we exclude  object 11, we obtain $r_p$=-0.64. 
Similarly, elliptical galaxies with larger $n$ reside in regions of higher
cluster surface density. Here, the correlation is stronger, $r_p$=0.63, and
$r_p$=0.76 excluding object 11. We can obtain similar results using the,
perhaps more meaningful, central galaxy concentration as defined by Trujillo,
Graham, \& Caon (2001):

\begin{equation}
C_{r_e}(\alpha)=\frac{\sum_{i,j\in E(\alpha r_e)}I_{ij}}{\sum_{i,j\in E(r_e)}I_{ij}}.
\end{equation}

Here, $E(r_e)$ is the isophote which encloses half of the total light of the
galaxy, and $E(\alpha r_e)$ is the isophote at a radius $\alpha(<1)$ times
$r_e$. Specifically, for a S\'ersic law one obtains that  $C_{r_e}(\alpha)$
increases monotonically with $n$

\begin{equation}
C_{r_e}(\alpha)=\frac{\gamma(2n,k\alpha^{1/n})}{\gamma(2n,k)}, 
\end{equation}

with $k=2n-0.324$ and where $\gamma(a,x)$ is the incomplete gamma function
(Abramowitz \& Stegun 1964). Figure 10 shows the relation between the central
galaxy concentration of each galaxy and a)  the (projected) distance from the
brightest central cluster member and b) the local cluster surface  density with
$\alpha=0.3$. The values for the Pearson's coefficient are similar than
obtained using $n$:  $r_p$=-0.45 and -0.56 (excluding object 11) for the
correlation between $C_{r_e}$ and distance, and $r_p$=0.68 ($r_p$=0.74 removing
object 11) for the correlation between $C_{r_e}$ and  the local cluster
surface  density.

The dispersion presents in  these relations might be inflated by the fact that
both the central distance and the local surface density are ``projected''
parameters. If the relation that we have found for A2443 is shown to hold for
other clusters, it means that the qualitative morphology-density relation noted
by previous authors (e.g.\ Dressler 1980; Dressler et al. 1997; Fasano et al.
2000) extends further and can be placed on a quantitative basis, such that the
detailed structure of the individual galaxies (beyond the broad
elliptical/spiral distinction) is related to their immediate
environment/density. The brightest cluster galaxies, which reside at the dense
centers of galaxy clusters, are known to contain multiple nuclei (Postman \&
Lauer 1995) from mergers. They also posess high values of $n$, or are better
described by power laws (Graham et al.\ 1996). What we may well be observing
here is an extension of this behavior to less massive systems and lower cluster
densities.

Strom \& Strom (1978) noted that the structural properties of Elliptical
galaxies are related to the dynamical properties of their parent clusters. By
tidal friction and high-speed impulse encounters they were able to explain why
characteristic sizes of the galaxies decrease by a factor of 1.5 or more in the
denser regions of the clusters. But tidal stripping seems not be the answer to
explain the previously shown correlations. In fact, galaxy central concentration
is an observed core property due to it is defined in the inner part (within the
effective radius) of the object. ``Encounters which lead to the stripping of
halo stars should not affect the distribution of stars in the tightly bound
cores of the galaxies'' (Strom \& Strom 1978). Also, in simulations, ``models
whose initial surface densities are described by an $r^{1/4}$ (de Vaucoleurs)
law are still well fitted by such a law after collisions in which the target
galaxy loses up to 40 percent of its mass'' (Aguilar \& White 1986).  Moreover,
``the $r^{1/n}$ models are stable to radial and non radial perturbations''
(Ciotti 1991). These statements seem to indicate that the global structure
(index $n$ or central concentration) of the galaxies must be only conditioned
by mechanisms which act over the complete structure (not only in the outer
parts). 

Against an explanation based on tidal stripping exists another evidence:
Graham, Trujillo \& Caon (2001) have simulated the loss of stars in the outer
parts of the galaxies by truncating $r^{1/n}$ profiles and remeasuring the
actual effective radius and, consequently, $C_{r_e}$. By doing this they find
that truncated galaxies tend to be less concentrated  than galaxies which
extend to infinity.

Favoring that what we are seeing is an effect of mechanisms which act over the
global structure are: a) mergers tend to increase the concentration of the
galaxies. As an example, numerical simulations (White 1983; Barnes 1990; Barnes
1992) have supported the hypothesis that mergers between typical disk galaxies
($n$=1) produce remnants with the overall morphology and structural
parameters of elliptical galaxies ($n$=4), b) the concentration (or index $n$)
correlates with global structure parameters of the galaxies such as total
luminosity, or equivalently mass.  (Caon et al. 1993; Young \& Currie 1994;
Jerjen \& Binggeli 1997).

Reasons exposed before suggest that  the increase in the central galaxy
concentration (that is, the larger values of $n$) as a function of the cluster
surface density may be a consequence of mergers, but also, an alternative to
this, may be that the galaxies with larger $n$, that is, larger central
concentration, may have formed from primordial fluctuations of greater
amplitude during a subclustering process. Theories of gravitational collapse
and galaxy formation must be able to distinguish between these two
alternatives.

\section{Summary}

We have presented broad-band $B$, $V$, $R$, and $I$ photometry of 121 galaxies
brighter than I=23 in the inner region of Abell Cluster 2443. We have modeled
the surface-brightness profiles from 33  of these objects and have obtained
reliable quantitative morphological classifications for 25 objects down to
$I=19$ mag, the remaining 8 being morphologically peculiar. Our structural
analysis procedure has been tested through Monte Carlo simulations.  We have
overcome two systematic problems that usually effect such fitting algorithms in
the past. Firstly, we do not evaluate the ellipticity as an external parameter
measured from the ellipticity of the outer isophotes; instead, we have taken
into account that the seeing-affected surface-brightness profile is a function
of the intrinsic ellipticity of the objects. Secondly, we have avoided the
numerical problems associated with the use of fast Fourier transforms to
compute the convolution between the surface brightness profile models and the
PSF of the images  when dealing with very steep intensity profiles. We have
determined the seeing-corrected central galaxy concentration (related to the
S\'ersic index). We reveal correlations between this parameter and the
projected  distance to the brightest cluster member and also with the projected
cluster surface density. This is possibly due to a larger number of mergers in
the high-density zones of the cluster. It is hoped trough the study of
more clusters that one will be able to determine the strength and possible
universality of this relationship, and learn important clues to the nature of
galaxy formation.

\acknowledgments

JALA was supported by grant 20-56888.99 from the Schweizerischer Nationalfonds.
Nordic Optical Telescope is operated on the island of La Palma  jointly by
Denmark, Finland, Iceland, Norway, and Sweden, in the Spanish Observatorio del
Roque de los Muchachos of the  Instituto de Astrofisica de Canarias.

We are  indebted to Alister W. Graham who kindly proof-read versions of this
manuscript. The authors are grateful to the anonymous referee for the valuable
refereeing that helped us to improve the rigor and clarity of this paper.

\clearpage

\figcaption[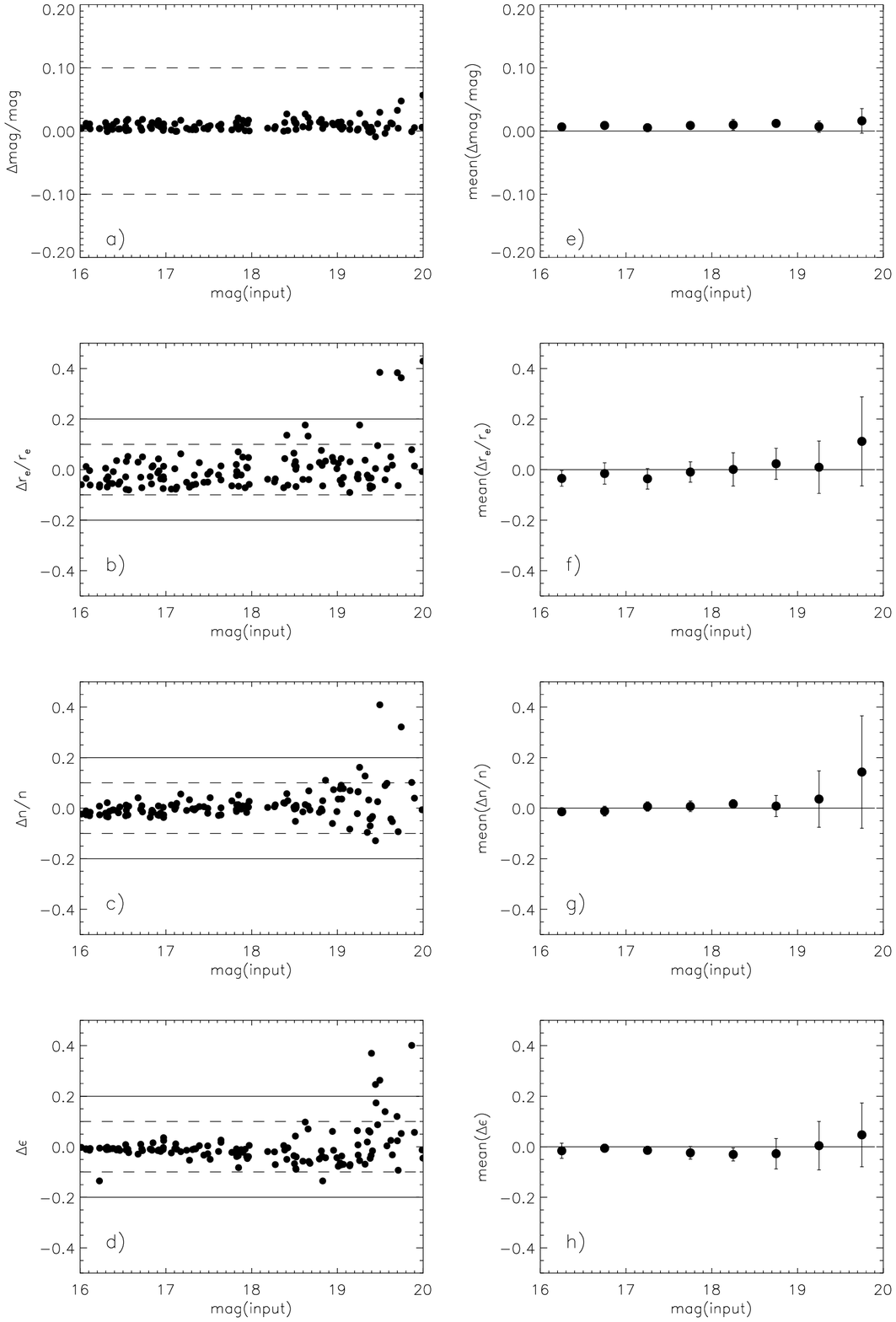]{$Left$ $from$ $the$ $top$ $downwards$: a) Relative error in the total recovery magnitude versus
the total input magnitude. b) Relative error of the output 
effective seeing-deconvolved radius versus the total input magnitude. c) Relative error of the output S\'ersic index versus the total input 
magnitude. d) The difference $\Delta\epsilon=\epsilon(output)-\epsilon(input)$ between measured and
input ellipticity versus the total input magnitude. $Right$ $from$ $the$
$top$ $downwards$: Mean
relative errors of the quantities shown in the left column with 1 $\sigma$
error bars.\label{f1}}

\figcaption[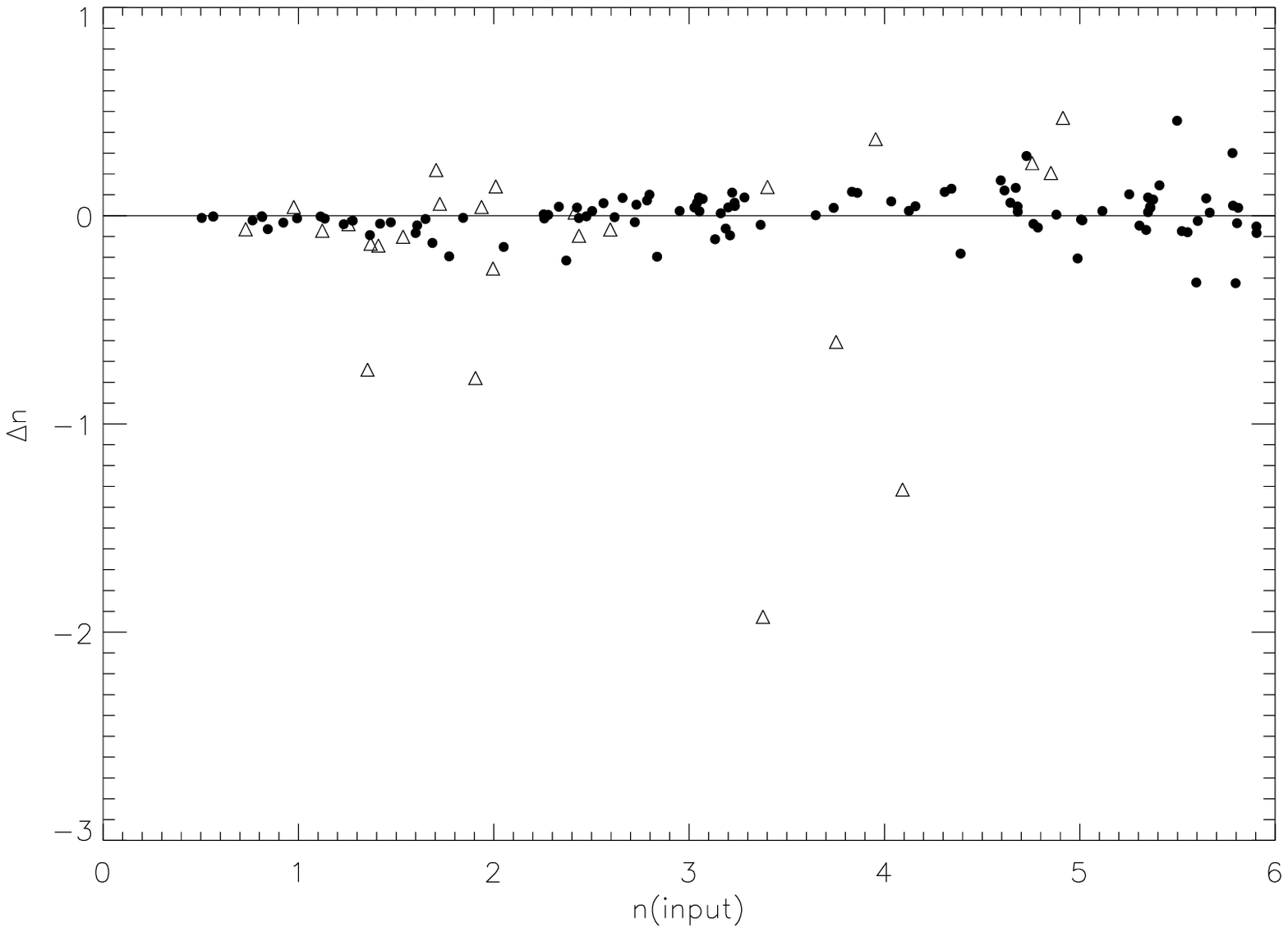]{Difference, $\Delta n$, between measured and input S\'ersic
indices versus input S\'ersic index. Two different magnitude intervals 
($I<19.5$ mag [solid circles], $I>19.5$ mag [open triangles]) are shown.\label{f2}}

\figcaption[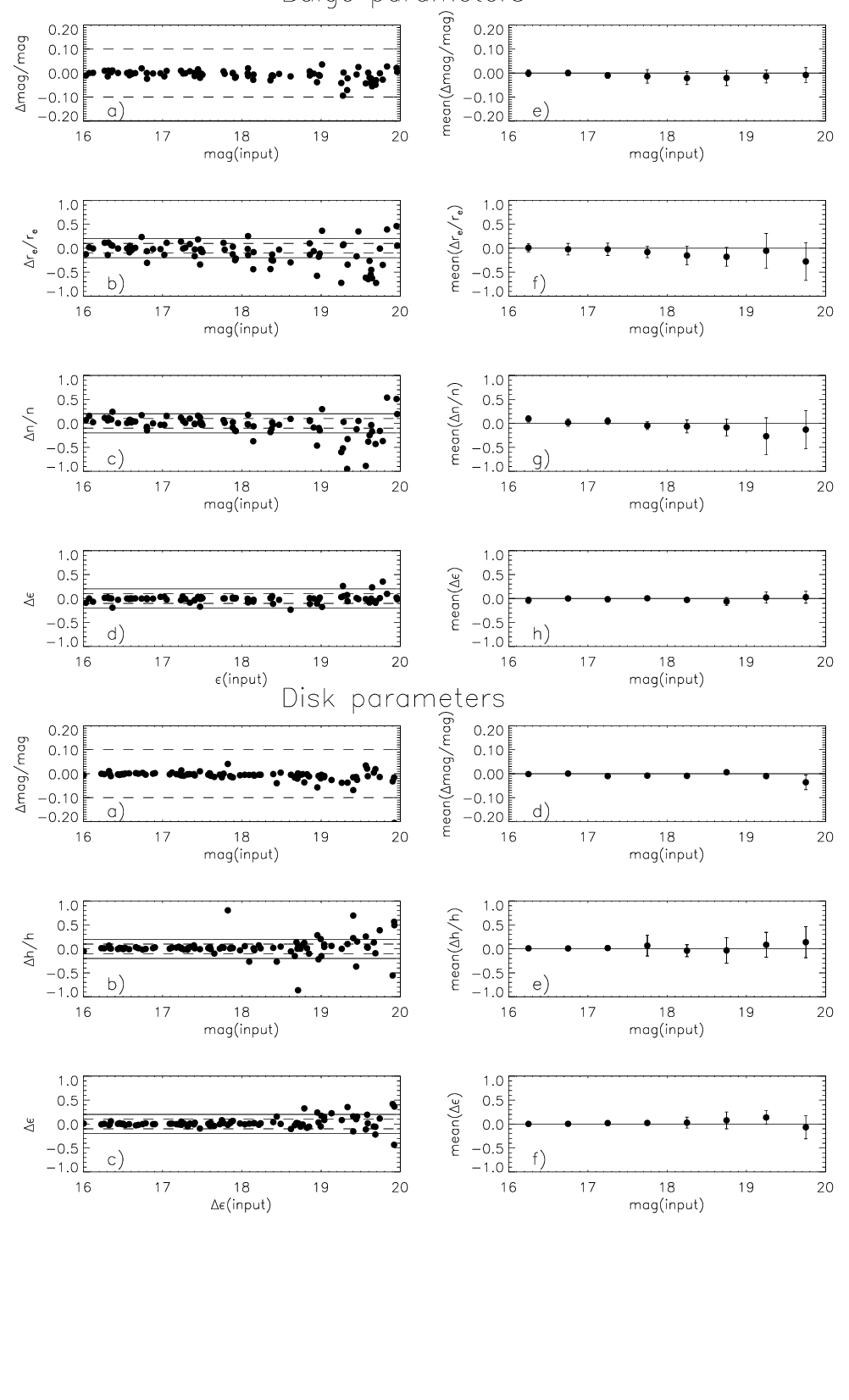]{$Upper$ $panel$: The recovery bulge parameters are
shown. $Left$ $from$ $the$ $top$ $downwards$: a) Relative error of the total 
recovery magnitude versus
the total input magnitude. b) Relative error of the output 
effective seeing-deconvolved radius versus the total input magnitude. c) Relative
 error of the output S\'ersic index versus the total input 
magnitude. d) The difference $\Delta\epsilon=\epsilon_{\rm b}(output)-\epsilon_{\rm b}(input)$ between measured and
input bulge ellipticity versus the total input magnitude. $Right$ $from$ $the$
 $top$ $downwards$: Mean
relative errors in the quantities shown in the left column with 1 $\sigma$
error bars.
$Lower$ $panel$: The recovery disk parameters are
shown. $Left$ $from$ $the$
 $top$ $downwards$: a) Relative error of the total recovery magnitude versus
the total input magnitude. b) Relative error of the output 
seeing-deconvolved scale length versus the total input magnitude. c) The difference
 $d\epsilon=\epsilon_{\rm d}(output)-\epsilon_{\rm d}(input)$ between measured and
input disk ellipticity versus the total input magnitude. $Right$ $from$ $the$
 $top$ $downwards$: Mean
relative errors of the quantities shown in the left column with 1 $\sigma$
error bars.
\label{f3}}

\figcaption[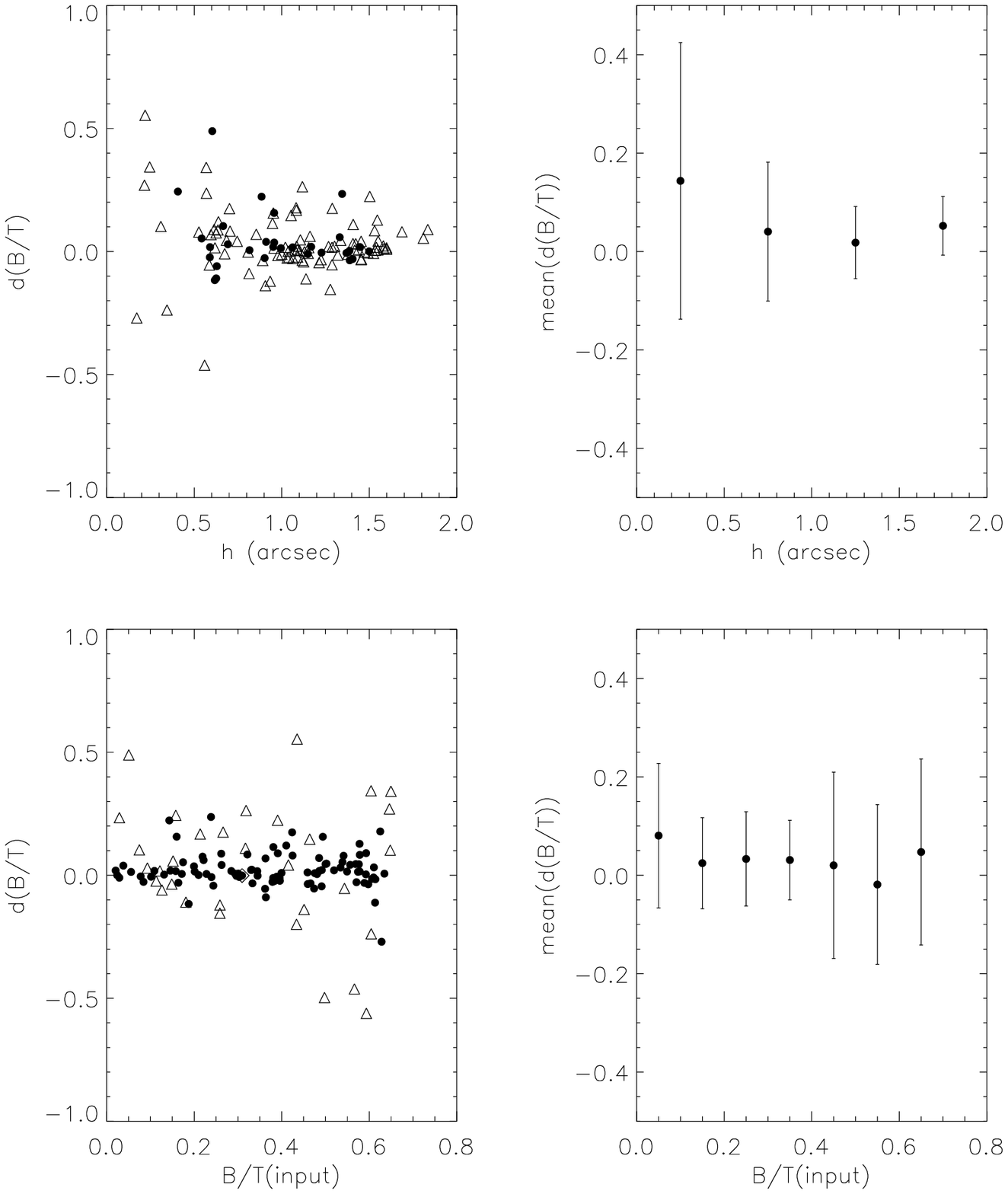]{$Top$ $left$: The difference, $d(B/T)$, between measured
(output) and input bulge-to-total luminosity ratios versus seeing-deconvolved
(i.e.\ model) scale length, $h$,  for two different magnitude intervals
($I<18.5$ mag [solid circles], $I>18.5$ mag  [open triangles]). $Top$ $right$:
Mean difference, $mean$ $d(B/T)$, between measured and input bulge-to-total
luminosity versus model  scale length with 1 $\sigma$ error bars. $Bottom$
$left$: The difference, $d(B/T)$, between measured and input $B/T$ versus the
$B/T$(input) for two different magnitude intervals  ($I<18.5$ mag [solid
circles], $I>18.5$ mag [open triangles]). $Bottom$ $right$: Mean difference,
$mean$ $d(B/T)$, between measured and input $B/T$ versus $B/T$(input) with 1
$\sigma$ error bars.\label{f4}}

\figcaption[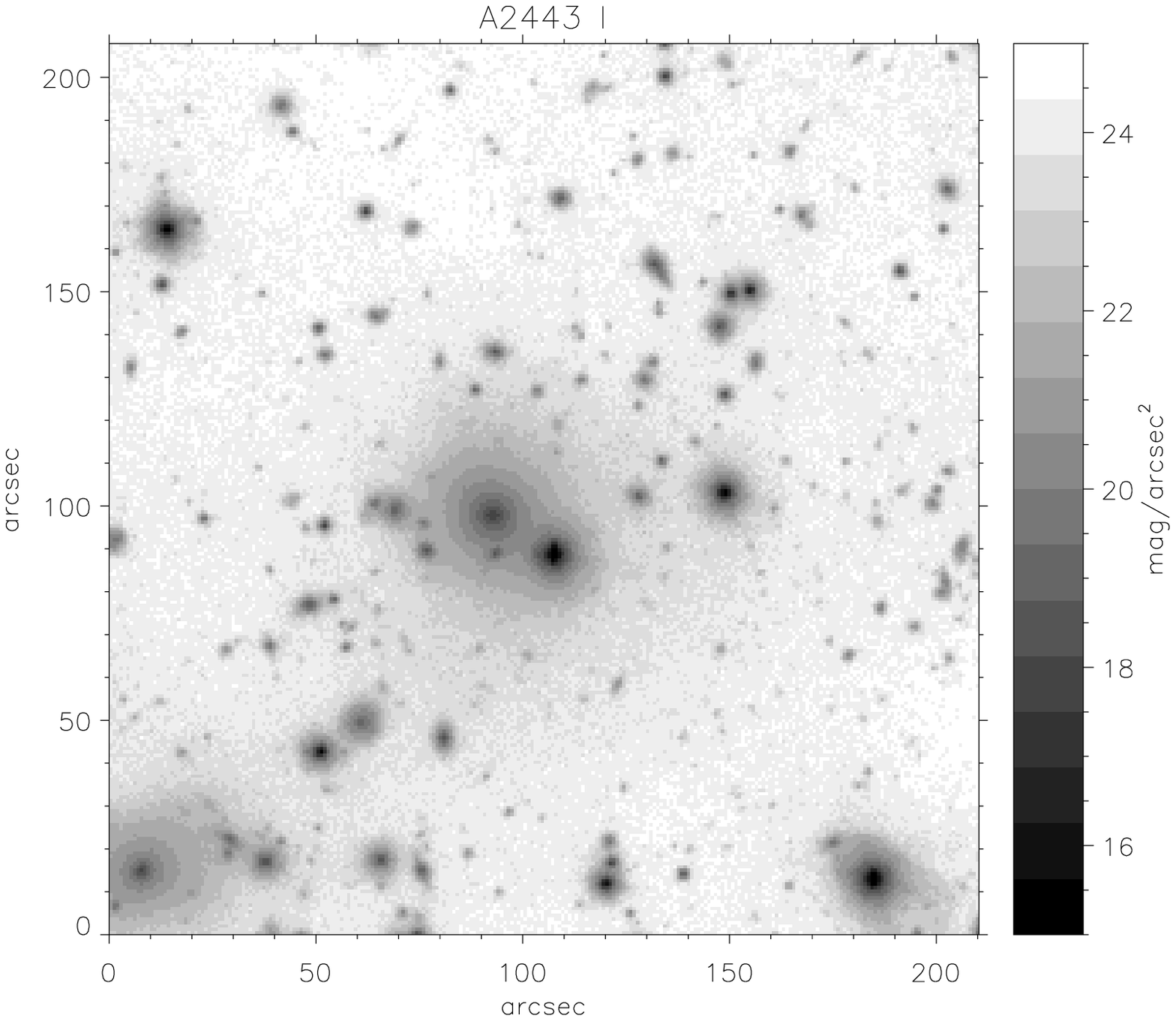]{The $I$-band image of the central part of the cluster A2443
observed at the NOT.\label{f5}}

\figcaption[Trujillo.fig6.ps]{Galaxies of the cluster A2443 for which   structural
parameters have been estimated.  From left to right is represented a gray-scale
image of the $I$ band data (surface brightness isocontours are shown to $\mu_I$=23
mag arcsec$^{-2}$ with steps of 1 mag arcsec$^{-2}$),  the surface brightness profile,
the ellipticity, and the residuals of the fit. Superimposed on the surface
brightness profile data are the intrinsic profile (dashed line) and the
convolution of this model profile (solid line) to match the data. The solid
lines in the ellipticity plots show the fit to the ellipticities using our
algorithm. Intrinsic ellipticities (i.e.\ not seeing convolved) of the galaxies
can be obtained by extrapolating to infinity the solid lines. \label{f6}} 

\figcaption[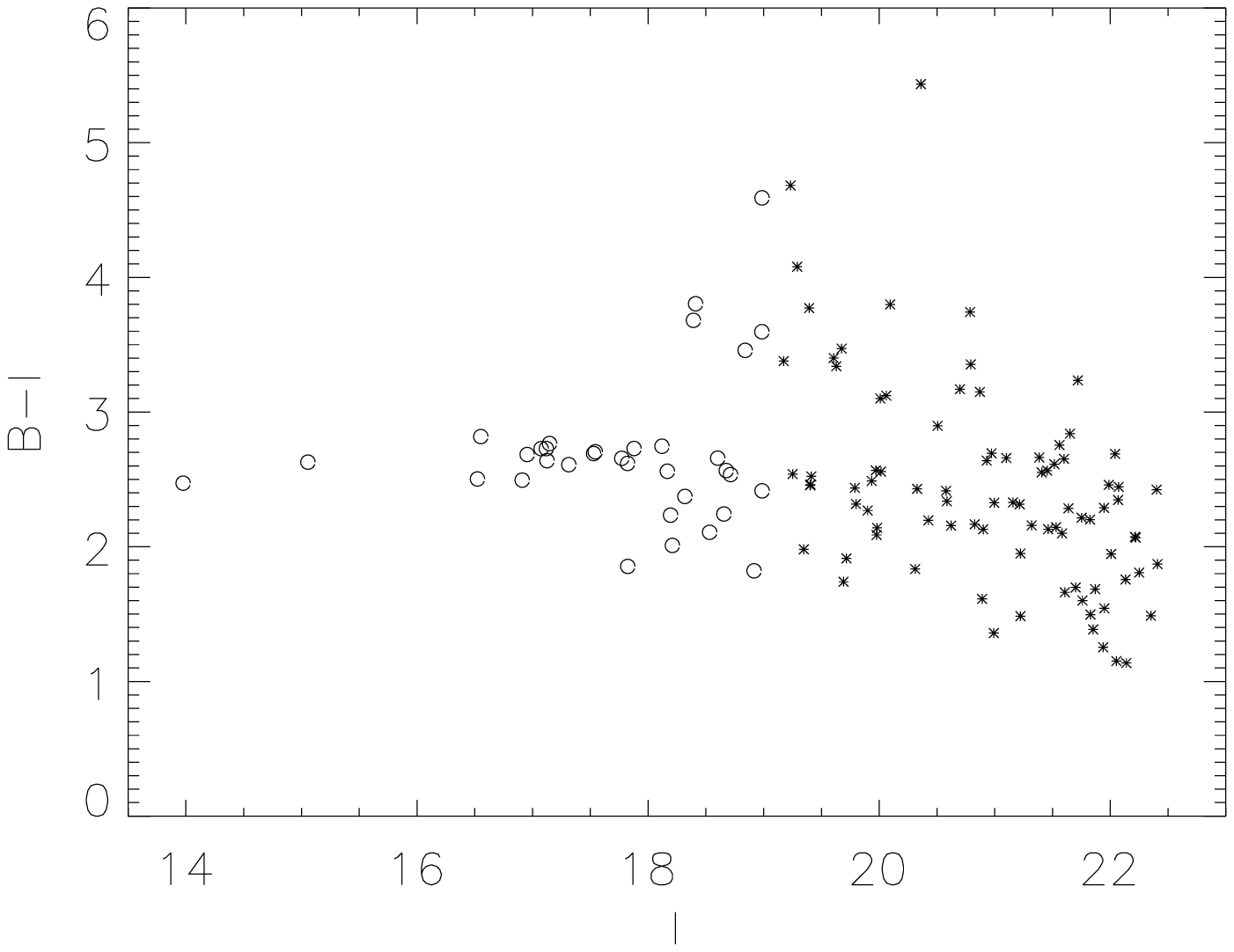]{Color--magnitude diagram for the central region
of A2443. Galaxies with $I<19$ mag (open symbols) were analyzed (see
text).\label{f7}}

\figcaption[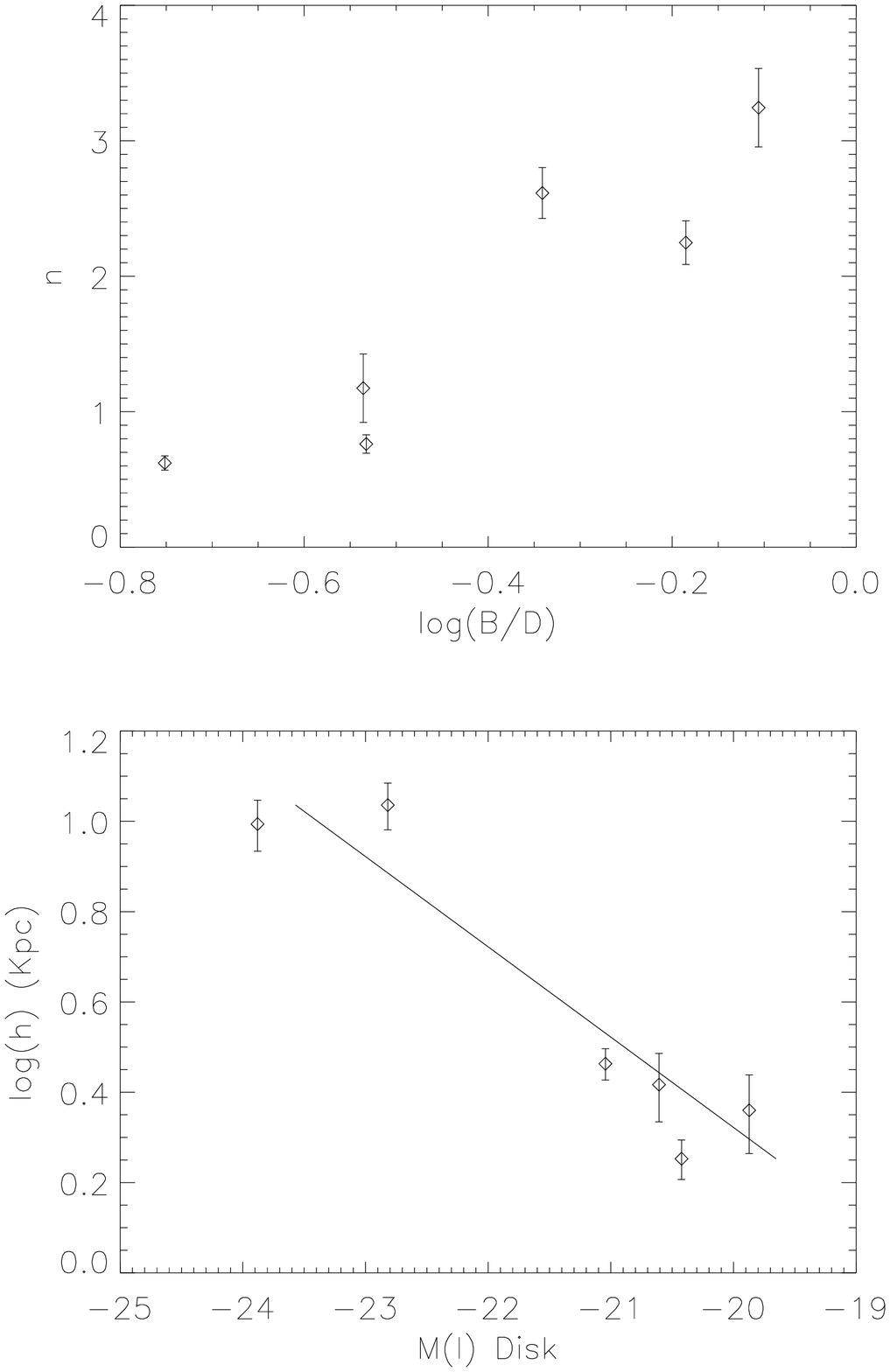]{$Top$: $n$ versus bulge-to-disk ratio for
 galaxies with $B/T<0.6$ $Bottom$: Relation between disk size and
luminosity for galaxies with $B/T<0.6$.\label{f8}}

 \figcaption[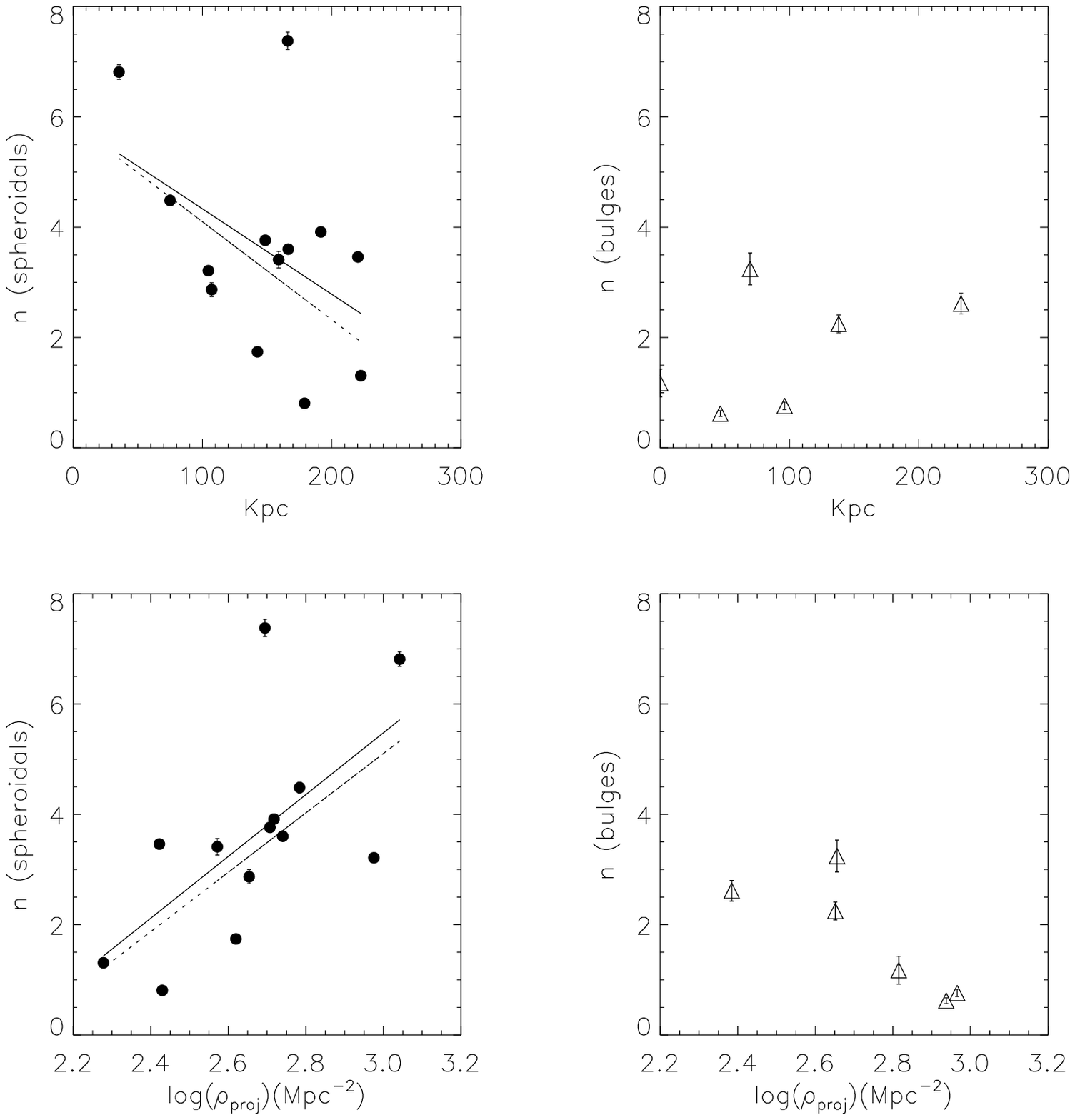]{$Top$ $left$: The index $n$
versus the distance to the brightest cluster member. The values of $n$ shown 
correspond to the galaxies referred as ``ellipticals''. $Top$ $right$: The
index $n$ versus the distance to the brightest cluster member. The values of
$n$ presented  correspond to the bulge component of the galaxies with $B/T <
0.6$. $Bottom$ $left$: The index $n$ versus the local surface density. The
values  of $n$ shown correspond to the galaxies referred to as ``ellipticals''.
$Bottom$ $right$: The index $n$ versus the local surface density. The values of
$n$ presented here correspond to the bulge component of the galaxies with $B/T
< 0.6$. Regression lines to all the points (solid lines) and excluding object 11
(dashed lines) are plotted.\label{f9}} 

 \figcaption[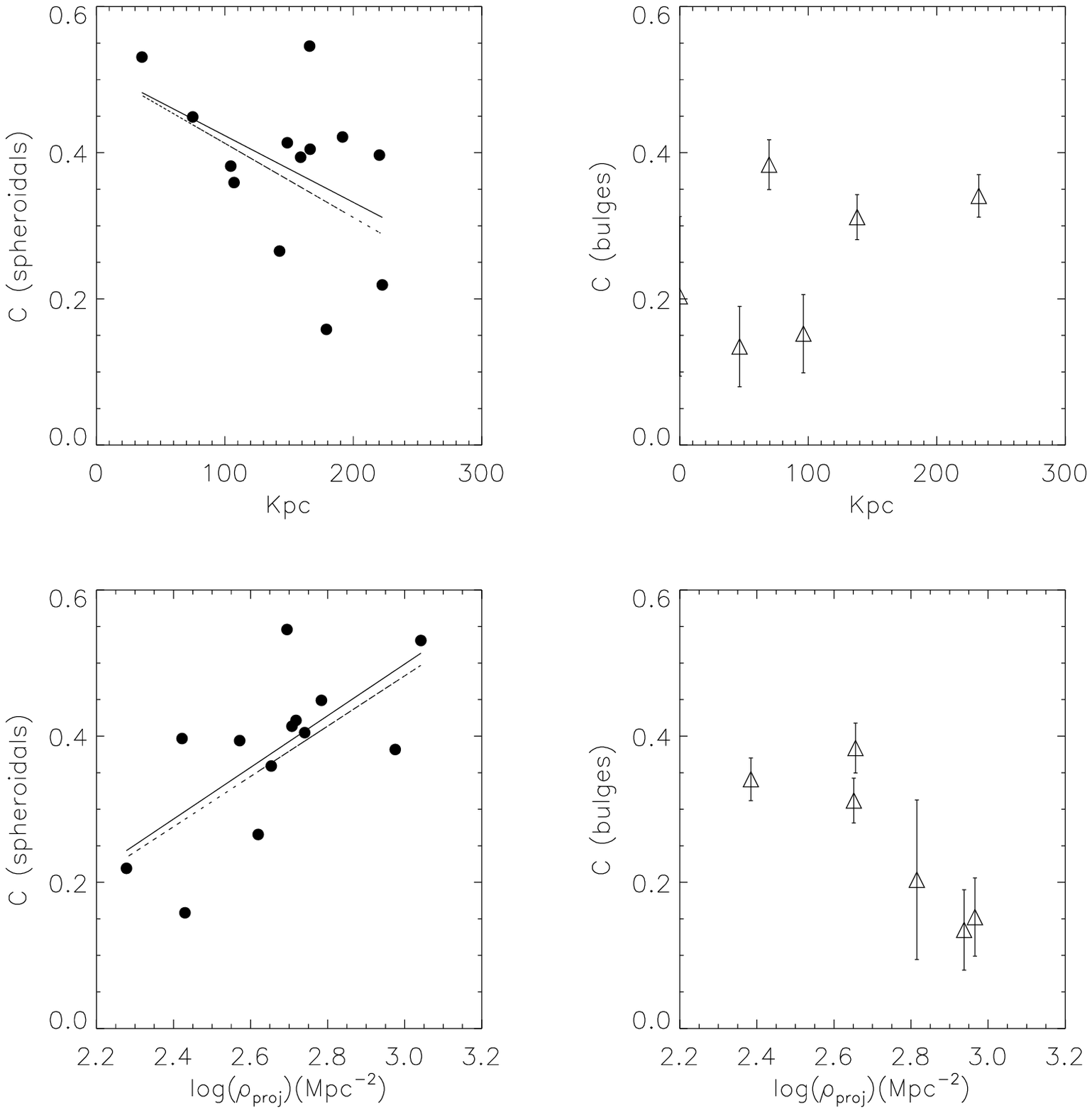]{$Top$ $left$: The central galaxy concentration
versus the distance to the brightest cluster member. The values of C shown 
correspond to the galaxies referred as ``ellipticals''. $Top$ $right$: C versus
the distance to the brightest cluster member. The values of C presented
correspond to the bulge component of the galaxies with $B/T < 0.6$. $Bottom$
$left$: C versus the local surface density. The values  of C shown correspond
to the galaxies referred to as ``ellipticals''. $Bottom$ $right$: C versus the
local surface density. The values of C presented here correspond to the bulge
component of the galaxies with $B/T < 0.6$. All the values were established
with $\alpha=0.3$. Regression lines to all the points (solid lines) and
excluding object 11 (dashed lines) are plotted.\label{f10}}

\clearpage

\begin{deluxetable}{ccccccccc}
\tabletypesize{\scriptsize}
\tablecaption{Mean relative errors for the bulge parameters}
\tablewidth{0pt}
\tablehead{\colhead{mag} & \colhead{16.25}& \colhead{16.75}& \colhead{17.25}& \colhead{17.75}& \colhead{18.25}& \colhead{18.75}& \colhead{19.25}& \colhead{19.75} }
\startdata
mean($\frac{\Delta mag}{mag}$)\tablenotemark{a} & 0.6$\pm$0.4 & 0.9$\pm$0.5 & 0.5$\pm$0.5 & 0.9 $\pm$ 0.6 & 1.0 $\pm$0.8 &
1.0$\pm$0.6 & 0.7$\pm$0.9 &1.6$\pm$1.9 \\
mean($\frac{\Delta r_{\rm e}}{r_{\rm e}}$)\tablenotemark{a} & --3.5$\pm$3.1 & --1.5$\pm$4.2 & --3.6$\pm$4.1 & --0.9$\pm$4.0 & 0.1$\pm$6.6 &2.3$\pm$6.1 & 0.9$\pm$1.0 &
11.2$\pm$17.6 \\
mean($\frac{\Delta n}{n}$)\tablenotemark{a} & --1.4$\pm$1.4 & --1.2$\pm$1.9 & 0.7$\pm$1.9 & 0.7$\pm$2.1 & 1.7$\pm$1.7 & 0.8$\pm$4.2 & 3.6$\pm$11.2 & 14.3$\pm$22.3\\ 
mean($\epsilon_{\rm out}-\epsilon_{\rm in}$)&  --0.02$\pm$0.03 &--0.01$\pm$0.02 & --0.01$\pm$0.02 & --0.02$\pm$0.02 & --0.03$\pm$0.02 & --0.03$\pm$0.06 & 0.04$\pm$0.09 & 0.05$\pm$0.13\\ 
 \enddata
\tablenotetext{a}{In \%.}
\end{deluxetable}

\clearpage

\begin{deluxetable}{ccccccccc}
\tabletypesize{\scriptsize}
\tablecaption{Mean relative errors for the bulge and disk parameters}
\tablewidth{0pt}
\tablehead{\colhead{mag} &  \colhead{16.25}& \colhead{16.75}& \colhead{17.25}& \colhead{17.75}& \colhead{18.25}& \colhead{18.75}& \colhead{19.25}& \colhead{19.75} }
\startdata
mean($\frac{\Delta mag}{mag}$)\tablenotemark{a,b} & --0.1 $\pm$1.4  &  0.01$\pm$1.1 & --1.0$\pm$1.2 & --1.4$\pm$2.8 &
--2.1$\pm$2.7 & --2.1$\pm$3.2 & --1.4$\pm$2.7 & --0.8$\pm$3.11 \\ 
mean($\frac{\Delta r_{e}}{r_{e}}$)\tablenotemark{a} &  0.1$\pm$8.7  & --2.0 $\pm$12.0 & --2.2$\pm$13.0 & --8.0$\pm$12.0 &
--21.8$\pm$42.06 & --31.1$\pm$35.61 & --16.5$\pm$46.9 & --8.4$\pm$ 30.6\\
mean($\frac{\Delta n}{n}$)\tablenotemark{a} &  9.6$\pm$6.8  &  1.7$\pm$7.8 & 4.8$\pm$6.7 & --4.6$\pm$8.2 &
--6.3$\pm$13.5 & --8.4$\pm$17.7 & --26.7$\pm$38.29 & --13.05$\pm$40.1 \\
mean($\epsilon_{\rm out}-\epsilon_{\rm in}$)\tablenotemark{b}& --0.06$\pm$0.05  &  --0.03$\pm$0.07 & --0.02$\pm$0.03 & --0.04$\pm$ 0.07&--0.03$\pm$0.15 & 0.01$\pm$0.16 & 0.11$\pm$0.20 & --0.03$\pm$0.10 \\
mean($\frac{\Delta mag}{mag}$)\tablenotemark{a,c}&  --0.1$\pm$0.2  &  0.01$\pm$0.01 & --0.9$\pm$1.0 & --0.8$\pm$0.8 &
--0.9$\pm$0.9 & 0.6$\pm$0.5 & --0.9$\pm$1.0 & 3.5$\pm$3.1 \\
mean($\frac{\Delta h}{h}$)\tablenotemark{a}&  1.1$\pm$2.6  &  9.7$\pm$1.7 & 1.7$\pm$2.4 & 6.7$\pm$21.5 &--3.7
$\pm$12.5 & --3.3$\pm$26.5 & 8.4$\pm$25.8 & 13.9$\pm$32.3 \\
mean($\epsilon_{\rm out}-\epsilon_{\rm in}$)\tablenotemark{c}&  0.01$\pm$0.02  &  0.01$\pm$0.04 & 0.02$\pm$0.05 & 0.02$\pm$0.03 &
0.03$\pm$0.11 & 0.08$\pm$0.18 & 0.14$\pm$0.14 & --0.07$\pm$0.24 \\
 \enddata
\tabletypesize{\normal}
\tablenotetext{a}{In \%.}
\tablenotetext{b}{Bulge parameter.}
\tablenotetext{c}{Disk parameter.}
\end{deluxetable}

\clearpage

\begin{deluxetable}{crrrrrr}
\tabletypesize{\scriptsize}
\tablecolumns{7} 
\tablewidth{0pc} 
\tablecaption{Broad-band photometry (Johnson filters) of the core
of Abell 2443 \label{tbl-3}}
\tablewidth{0pt}
\tablehead{
\colhead{ID} & \colhead{R.A. (2000)}   & \colhead{Dec. (2000)}   &
\colhead{$B$} &
\colhead{$V$}  & \colhead{$R$} & \colhead{$I$} }
\startdata
   1    &  22:26:07.92  &  17:21:24.9 & 16.44 & 15.17 & 14.83 & 13.97 \\
   2    &  22:26:04.61  &  17:21:55.4 & 19.02 & 17.69 & 17.35 & 16.52 \\
   3    &  22:26:01.86  &  17:22:35.3 & 23.23 & 23.42 & 19.59 & 17.42 \\
   4    &  22:26:02.38  &  17:22:17.8 & 19.37 & 17.85 & 17.27 & 16.55 \\ 
   5    &  22:26:01.65  &  17:22:53.0 & 20.72 & 19.87 & 19.23 & 18.16 \\
   6    &  22:26:02.72  &  17:22:26.3 & 20.22 & 18.77 & 18.28 & 17.52 \\
   7    &  22:26:02.21  &  17:22:46.9 & 17.68 & 16.27 & 15.86 & 15.05 \\
   8    &  22:26:01.76  &  17:19:42.8 & 19.67 & 19.62 & 19.58 & 17.82 \\
   9    &  22:26:02.76  &  17:20:04.4 & 20.22 & 19.32 & 18.87 & 18.21 \\
  10    &  22:26:02.42  &  17:21:50.8 & 19.63 & 18.21 & 17.80 & 16.95 \\
  11    &  22:26:02.26  &  17:21:41.2 & 20.42 & 19.01 & 18.59 & 17.77 \\
  12    &  22:26:01.59  &  17:21:45.5 & 23.37 & 21.72 & 20.41 & 19.29 \\
  13    &  22:26:01.61  &  17:20:52.2 & 25.79 & 23.05 & 22.18 & 20.36 \\
  14    &  22:26:01.57  &  17:20:23.0 & 24.28 & 23.33 & 22.79 & 22.21 \\
  16    &  22:26:14.36  &  17:22:37.2 & 23.39 & 22.60 & 22.27 & 21.70 \\
  17    &  22:26:13.98  &  17:20:52.9 & 23.35 & 23.20 & 22.55 & 21.75 \\
  18    &  22:26:13.88  &  17:21:47.5 & 22.62 & 21.80 & 21.19 & 20.42 \\
  19    &  22:26:13.90  &  17:21:26.6 & 24.01 & 23.12 & 22.42 & 21.45 \\
  20    &  22:26:13.77  &  17:20:15.8 & 23.00 & 21.90 & 20.79 & 19.60 \\
  21    &  22:26:13.71  &  17:20:43.5 & 22.53 & 21.28 & 20.78 & 19.97 \\
  22    &  22:26:13.70  &  17:20:24.5 & 23.92 & 22.63 & 22.20 & 21.63 \\
  23    &  22:26:13.60  &  17:20:51.6 & 22.22 & 20.97 & 20.51 & 19.78 \\
  25    &  22:26:13.28  &  17:22:43.1 & 23.95 & 22.99 & 22.37 & 21.40 \\
  26    &  22:26:13.18  &  17:19:38.8 & 22.21 & 20.55 & 19.53 & 18.41 \\
  27    &  22:26:12.98  &  17:21:09.7 & 20.86 & 19.37 & 18.94 & 18.11 \\
  28    &  22:26:12.56  &  17:22:35.4 & 21.86 & 20.47 & 19.89 & 19.40 \\
  29    &  22:26:12.74  &  17:20:12.8 & 22.58 & 21.14 & 20.13 & 18.98 \\
  30    &  22:26:12.72  &  17:20:32.5 & 24.73 & 23.23 & 22.73 & 22.04 \\
  31    &  22:26:12.57  &  17:20:11.1 & 23.32 & 22.38 & 21.77 & 20.99 \\
  32    &  22:26:12.48  &  17:21:44.6 & 21.32 & 20.25 & 19.95 & 19.34 \\
  33    &  22:26:12.56  &  17:21:19.5 & 24.44 & 23.10 & 22.70 & 21.98 \\
  34    &  22:26:12.51  &  17:22:51.6 & 24.41 & 22.94 & 20.73 & 22.06 \\
  35    &  22:26:11.93  &  17:20:47.1 & 19.80 & 18.35 & 17.94 & 17.07 \\
  36    &  22:26:11.44  &  17:21:40.8 & 23.59 & 22.25 & 21.95 & 21.46 \\
  37    &  22:26:11.47  &  17:19:46.4 & 24.14 & 22.65 & 21.77 & 20.79 \\
  38    &  22:26:11.08  &  17:21:52.9 & 21.24 & 19.91 & 19.51 & 18.71 \\
  39    &  22:26:10.96  &  17:20:31.9 & 19.84 & 18.39 & 17.95 & 17.11 \\
  40    &  22:26:10.88  &  17:22:06.2 & 21.26 & 19.98 & 19.42 & 18.60 \\
  41    &  22:26:10.92  &  17:21:06.2 & 23.57 & 22.05 & 21.66 & 20.93 \\
  42    &  22:26:10.80  &  17:22:38.4 & 22.06 & 20.94 & 20.67 & 19.97 \\
  43    &  22:26:10.80  &  17:21:04.6 & 24.02 & 22.22 & 21.96 & 20.87 \\
  44    &  22:26:10.53  &  17:21:24.8 & 20.44 & 19.00 & 18.59 & 17.82 \\
  45    &  22:26:10.46  &  17:22:04.6 & 21.79 & 20.41 & 20.02 & 19.25 \\
  46    &  22:26:10.49  &  17:20:58.8 & 23.55 & 22.67 & 22.39 & 21.86 \\
  47    &  22:26:10.37  &  17:21:37.8 & 22.11 & 20.91 & 20.51 & 19.79 \\
  48    &  22:26:10.39  &  17:20:23.6 & 21.24 & 19.85 & 19.50 & 18.67 \\
  49    &  22:26:10.11  &  17:20:49.5 & 20.89 & 19.86 & 19.45 & 18.65 \\
  50    &  22:26:10.23  &  17:22:50.4 & 21.43 & 20.71 & 20.39 & 19.69 \\
  51    &  22:26:10.34  &  17:22:10.8 & 24.23 & 22.82 & 22.56 & 21.94 \\
  52    &  22:26:10.10  &  17:21:04.5 & 22.42 & 21.10 & 20.76 & 19.93 \\
  53    &  22:26:09.91  &  17:21:14.8 & 22.16 & 20.91 & 22.19 & 19.89 \\
  54    &  22:26:09.92  &  17:21:29.3 & 22.55 & 21.10 & 20.46 & 19.17 \\
  55    &  22:26:10.38  &  17:20:47.7 & 22.96 & 20.69 & 20.39 & 19.62 \\
  56    &  22:26:09.37  &  17:21:09.8 & 23.19 & 22.60 & 22.13 & 21.93 \\
  57    &  22:26:09.38  &  17:20:32.6 & 23.20 & 22.44 & 22.35 & 22.05 \\
  58    &  22:26:09.38  &  17:19:46.5 & 23.48 & 22.65 & 21.98 & 21.15 \\
  59    &  22:26:09.29  &  17:22:09.2 & 24.49 & 22.91 & 22.30 & 21.65 \\
  61    &  22:26:07.35  &  17:21:40.5 & 19.76 & 18.22 & 17.78 & 17.12 \\
  62    &  22:26:08.09  &  17:21:52.6 & 20.69 & 19.45 & 19.09 & 18.31 \\
  63    &  22:26:08.00  &  17:21:47.9 & 19.40 & 18.11 & 17.70 & 16.91 \\
  64    &  22:26:08.71  &  17:19:38.1 & 23.89 & 21.96 & 21.18 & 20.09 \\
  65    &  22:26:08.25  &  17:20:50.8 & 20.60 & 19.14 & 19.00 & 17.87 \\
  66    &  22:26:09.05  &  17:21:51.6 & 23.49 & 22.25 & 22.29 & 21.94 \\
  68    &  22:26:08.18  &  17:19:41.9 & 21.93 & 20.56 & 20.26 & 19.41 \\
  69    &  22:26:08.18  &  17:19:55.0 & 23.83 & 23.47 & 23.02 & 22.35 \\
  70    &  22:26:08.41  &  17:19:48.5 & 23.86 & 22.97 & 24.71 & 20.69 \\
  71    &  22:26:07.88  &  17:19:54.8 & 22.92 & 21.49 & 21.08 & 20.58 \\
  72    &  22:26:07.83  &  17:22:32.9 & 23.16 & 21.59 & 20.95 & 19.39 \\
  73    &  22:26:07.69  &  17:22:08.9 & 23.17 & 22.69 & 21.95 & 21.22 \\
  74    &  22:26:07.48  &  17:22:53.3 & 20.42 & 19.36 & 18.98 & 18.19 \\
  75    &  22:26:07.46  &  17:19:34.9 & 20.64 & 19.66 & 19.32 & 18.53 \\
  76    &  22:26:07.43  &  17:20:44.1 & 23.47 & 22.80 & 22.15 & 21.31 \\
  77    &  22:26:07.32  &  17:21:24.3 & 19.91 & 18.32 & 17.93 & 17.14 \\
  78    &  22:26:07.01  &  17:21:52.5 & 24.28 & 23.02 & 22.78 & 22.22 \\
  79    &  22:26:07.09  &  17:19:40.8 & 23.95 & 23.85 & 23.26 & 22.00 \\
  80    &  22:26:06.94  &  17:19:38.6 & 23.91 & 21.34 & 20.48 & 19.23 \\
  81    &  22:26:06.76  &  17:19:39.6 & 21.85 & 20.57 & 20.20 & 19.40 \\
  82    &  22:26:06.47  &  17:22:07.9 & 20.24 & 18.76 & 18.37 & 17.54 \\
  83    &  22:26:06.51  &  17:19:53.9 & 23.14 & 21.57 & 20.73 & 19.67 \\
  84    &  22:26:06.40  &  17:21:52.0 & 24.02 & 22.91 & 22.40 & 21.82 \\
  85    &  22:26:08.81  &  17:20:45.4 & 22.29 & 20.75 & 20.06 & 18.83 \\
  86    &  22:26:06.56  &  17:22:02.3 & 23.57 & 21.62 & 20.93 & 18.98 \\
  87    &  22:26:06.13  &  17:21:58.1 & 22.35 & 21.70 & 21.63 & 20.99 \\
  88    &  22:26:06.17  &  17:22:00.0 & 24.95 & 23.52 & 23.12 & 21.71 \\
  89    &  22:26:06.22  &  17:19:45.8 & 22.75 & 21.52 & 21.14 & 20.32 \\
  90    &  22:26:05.81  &  17:22:17.1 & 20.73 & 19.81 & 19.52 & 18.91 \\
  91    &  22:26:05.88  &  17:21:52.2 & 24.12 & 22.57 & 23.04 & 21.51 \\
  92    &  22:26:05.82  &  17:21:28.0 & 23.67 & 22.84 & 22.65 & 21.53 \\
  93    &  22:26:05.85  &  17:20:31.3 & 23.02 & 22.41 & 21.95 & 20.89 \\
  94    &  22:26:05.75  &  17:22:27.4 & 22.57 & 21.17 & 20.81 & 20.01 \\
  95    &  22:26:05.69  &  17:21:16.5 & 23.53 & 22.46 & 21.99 & 21.21 \\
  96    &  22:26:05.73  &  17:21:44.6 & 24.31 & 23.31 & 22.80 & 21.55 \\
  97    &  22:26:05.75  &  17:20:01.2 & 22.12 & 21.09 & 20.79 & 19.98 \\
  98    &  22:26:05.73  &  17:19:37.7 & 22.99 & 21.81 & 21.41 & 20.82 \\
  99    &  22:26:05.23  &  17:20:55.7 & 23.11 & 21.44 & 20.92 & 20.00 \\
 100    &  22:26:05.18  &  17:21:50.5 & 23.40 & 21.87 & 21.23 & 20.50 \\
 101    &  22:26:05.21  &  17:21:27.5 & 24.05 & 23.72 & 27.27 & 22.25 \\
 102    &  22:26:04.94  &  17:22:51.6 & 22.70 & 22.51 & 21.87 & 21.22 \\
 103    &  22:26:04.79  &  17:20:32.2 & 24.82 & 23.29 & 22.68 & 22.40 \\
 104    &  22:26:04.37  &  17:21:36.2 & 19.92 & 18.52 & 18.14 & 17.31 \\
 105    &  22:26:04.36  &  17:22:31.8 & 23.27 & 23.01 & 22.69 & 22.13 \\
 106    &  22:26:04.41  &  17:20:40.1 & 23.88 & 23.04 & 22.50 & 22.13 \\
 107    &  22:26:04.10  &  17:22:37.8 & 22.77 & 21.80 & 21.46 & 20.62 \\
 108    &  22:26:03.63  &  17:20:53.8 & 23.26 & 22.37 & 22.03 & 21.60 \\
 109    &  22:26:05.00  &  17:21:03.1 & 24.51 & 23.13 & 22.34 & 22.07 \\
 113    &  22:26:02.70  &  17:22:14.3 & 23.18 & 21.25 & 20.81 & 20.06 \\
 114    &  22:26:15.27  &  17:19:59.6 & 22.50 & 22.51 & 22.08 & 20.89 \\
 116    &  22:26:03.20  &  17:21:20.8 & 24.52 & 23.09 & 22.43 & 20.78 \\
 117    &  22:26:15.17  &  17:20:31.1 & 22.14 & 22.51 & 21.65 & 20.31 \\
 118    &  22:26:02.91  &  17:22:17.4 & 23.96 & 22.48 & 22.19 & 21.75 \\
 119    &  22:26:02.84  &  17:21:41.7 & 24.25 & 23.47 & 22.69 & 21.59 \\
 120    &  22:26:02.75  &  17:20:57.1 & 21.40 & 20.12 & 19.65 & 18.98 \\
 121    &  22:26:02.62  &  17:21:40.2 & 24.28 & 22.88 & 22.51 & 22.40 \\
 122    &  22:26:02.64  &  17:20:38.5 & 23.32 & 22.70 & 22.04 & 21.82 \\
 123    &  22:26:02.53  &  17:21:30.4 & 22.99 & 22.15 & 21.38 & 20.57 \\
 124    &  22:26:02.07  &  17:20:14.8 & 23.66 & 22.35 & 22.63 & 20.97 \\
 125    &  22:26:01.86  &  17:22:08.9 & 24.04 & 22.62 & 22.54 & 21.38 \\
 126    &  22:26:01.76  &  17:21:04.2 & 23.68 & 22.97 & 22.58 & 21.58 \\
 127    &  22:26:14.74  &  17:21:02.3 & 21.63 & 20.84 & 20.38 & 19.71 \\
 128    &  22:26:14.75  &  17:20:58.9 & 23.75 & 22.77 & 22.06 & 21.10 \\
 129    &  22:26:14.41  &  17:22:15.3 & 22.07 & 20.53 & 19.42 & 18.39 \\
\enddata

\end{deluxetable}
\clearpage

\begin{deluxetable}{ccccccccc}
\tabletypesize{\scriptsize} 
\tablecolumns{9} 
\tablewidth{0pc} 
\tablecaption{Parameters of the bulge and disk decomposition for objects with
$B/T<0.6$} 
\tablehead{ 
\colhead{ID}    & \colhead{$I_{\rm e}$}& \colhead{$r_{\rm e}$}& \colhead{$n$}& \colhead{$\epsilon_{\rm b}$}& \colhead{$I_{0}$}& \colhead{$h$}& \colhead{$\epsilon_{\rm d}$} & \colhead{$B/T$}\\
 & (mag arcsec$^{-2}$) & (arcsec) & & & (mag arcsec$^{-2}$) & (arcsec) & &} 
\startdata 
1  &     19.46 $\pm$      0.05  &     1.73 $\pm$     0.15&
 1.17 $\pm$    0.05  &   0.05  &     19.53$\pm$
 0.19  &     4.94  $\pm$    0.64  &    0.12 & 0.23 \\
7  &     19.92  $\pm$    0.08  &     1.43 $\pm$   0.08&
 2.61 $\pm$    0.08  &    0.12 &      20.71$\pm$
 0.12  &     5.44 $\pm$     0.64 &     0.19&0.31\\
39  &     18.71  $\pm$     0.40  &    0.40  $\pm$    0.11&
 2.25 $\pm$     0.66  &    0.25   &    20.06$\pm$
 0.39  &     1.31$\pm$      0.23 &   0.01& 0.39 \\
63  &     17.88   $\pm$    0.24  &    0.22  $\pm$   0.02&
0.62  $\pm$    0.75 &     0.20  &     19.73$\pm$
 0.11  &     1.45  $\pm$    0.12 &     0.11& 0.15\\
65  &     20.01   $\pm$    0.3  &    0.48  $\pm$   0.10&
 3.24  $\pm$    0.82  &    0.16 &      20.31$\pm$
 0.33  &     1.15 $\pm$     0.23 &     0.19&0.44\\
82  &     18.16  $\pm$     0.25  &    0.31 $\pm$    0.03&
      0.76 $\pm$     0.55  &    0.56  &     18.91$\pm$
       0.21  &    0.90 $\pm$    0.09 &     0.37&0.23\\
\enddata 
\end{deluxetable}
\clearpage

\begin{deluxetable}{ccccc} 
\tablecolumns{5} 
\tablewidth{0pc}
\tablecaption{Parameters of  objects with $B/T>0.6$ (elliptical objects)} 
\tablehead{ 
\colhead{ID}    & \colhead{$I_{\rm e}$} & \colhead{$r_{\rm e}$} & \colhead{$n$} & \colhead{$\epsilon_{\rm b}$}\\
& (mag arcsec$^{-2})$ & (arcsec) & &}
\startdata 
4   &    21.60  $\pm$     0.04  &     2.86  $\pm$   0.07&
       3.91 $\pm$    0.06 &     0.12 \\
6   &   27.36   $\pm$   0.24  &     66.37  $\pm$     10.16&
       10.05 $\pm$     0.33 &     0.53 \\
9  &     23.14  $\pm$     0.75   &    3.20   $\pm$    1.99&
       3.46 $\pm$      1.16 &     0.38 \\
10  &     21.12 $\pm$    0.05 &      1.72 $\pm$    0.05&
       3.60 $\pm$    0.07 &    0.03 \\            
11  &     22.09  $\pm$     0.06 &      2.26 $\pm$    0.08&
       7.38 $\pm$     0.19 &     0.43 \\                            
26   &    23.23  $\pm$     0.10 &      2.53 $\pm$   0.14&
       8.82 $\pm$     0.36 &     0.17 \\      
27  &     20.91  $\pm$    0.01 &     0.97 $\pm$  0.01& 
       3.76 $\pm$    0.02 &     0.21 \\           
29  &     22.68  $\pm$     0.47 &      1.53 $\pm$   0.46&
       5.83 $\pm$      1.08 &     0.24 \\            
38  &     21.81  $\pm$     0.49 &      1.29 $\pm$    0.41&
       2.87 $\pm$     0.81 &     0.27 \\                 
44  &     21.35  $\pm$     0.18 &      1.62 $\pm$   0.16&
       4.49 $\pm$     0.30 &     0.33\\                 
48  &     20.98  $\pm$     0.22  &     1.02 $\pm$   0.12&
       1.74 $\pm$     0.29 &     0.31 \\            
61  &     21.72  $\pm$     0.57 &      1.86 $\pm$     0.71&
       6.81 $\pm$      1.29 &     0.14 \\                            
74  &     21.39  $\pm$     0.11  &     1.74 $\pm$     0.12&
      0.81 $\pm$    0.08 &     0.24 \\     
75  &     21.73  $\pm$     0.17  &     1.79 $\pm$     0.18&
       1.31 $\pm$     0.17 &     0.52 \\   
85  &     15.57 $\pm$      0.60  &   0.04 $\pm$    0.02&
       8.87 $\pm$      7.84 &     0.10 \\    
86  &     15.02 $\pm$      0.08  &   0.03 $\pm$   0.01&
       12.29 $\pm$      2.97 &   0.01 \\ 
104 &      20.84 $\pm$      0.16&       1.66 $\pm$    0.15&
       3.21 $\pm$     0.22 &     0.32 \\
120  &     22.27 $\pm$      0.20 &      1.41 $\pm$   0.16&
       3.41 $\pm$     0.34 &     0.25 \\      
129  &     22.29 $\pm$      0.04 &      1.58 $\pm$  0.04&
       4.79 $\pm$     0.16 &     0.10 \\          
\enddata 
\end{deluxetable} 
\end{document}